\documentclass[prd,twocolumn,groupedaddress,showpacs,nofootinbib]{revtex4}
\usepackage{graphicx}
\usepackage{dcolumn}
\usepackage{bm}
\usepackage{amssymb}
\usepackage{mathrsfs}
\usepackage{amsmath}
\usepackage{epsfig}
\usepackage[dvips]{color}
\usepackage{hhline}

\begin{document}

\title{Connections between the Seesaw and Dark Matter Searches}

\author{Adisorn Adulpravitchai}
\email{adisorn.adulpravitchai@mpi-hd.mpg.de}

\author{Pei-Hong Gu}
\email{peihong.gu@mpi-hd.mpg.de}

\author{Manfred Lindner}
\email{manfred.lindner@mpi-hd.mpg.de}

\affiliation{Max-Planck-Institut f\"{u}r Kernphysik, Saupfercheckweg
1, 69117 Heidelberg, Germany}

\begin{abstract}

In some dark matter models, the coupling of the dark matter particle
to the standard model Higgs determines the dark matter relic density
while it is also consistent with dark matter direct detection
experiments. On the other hand, the seesaw for generating the
neutrino masses probably arises from a spontaneous symmetry breaking
of global lepton number. The dark matter particle thus can
significantly annihilate into massless Majorons when the lepton
number breaking scale and hence the seesaw scale is near the
electroweak scale. This leads to an interesting interplay between
neutrino physics and dark matter physics and the annihilation mode
has an interesting implication on dark matter searches.

\end{abstract}

\pacs{95.35.+d, 14.60.Pq, 14.80.Bn}

\maketitle

\section{Introduction}

The existence of non-baryonic cold dark matter \cite{dunkley2008}
indicates the necessity of supplementing the $SU(3)_c^{}\times
SU(2)_{L}^{}\times U(1)_{Y}^{}$ standard model (SM) with new
ingredients. This has led to many interesting dark matter models.
For example, the dark matter particle may be a scalar field
\cite{sz1985,bpv2000,hllt2009,gms2010}. In this case, the dark
matter scalar can have a quartic coupling to the SM Higgs doublet.
Through the t-channel exchange of the Higgs boson, there will be
elastic scattering of the dark matter scalar by nucleons. This opens
a window for dark matter direct detection experiments
\cite{DMreviews}. In the case where the dark matter particle is not
a scalar, but a vector or a fermion, it can indirectly couple to the
SM Higgs. It does this by coupling directly to a non-SM Higgs, which
mixes with the SM Higgs in presence of the Higgs portal
\cite{hambye2009}. Direct detection experiments give strict upper
bounds on the dark matter-nucleon cross section. It should also be
noted that a dark-matter-Higgs coupling may also be responsible for
and hence constrained by the relic density of dark matter.

On the other hand, neutrino oscillation experiments prove that
neutrinos have masses and mixings \cite{stv2008} which also requires
new physics beyond the SM. The cosmological bound shows that
neutrino masses should be in the sub-eV range \cite{dunkley2008}.
The small neutrino masses can be naturally explained in the seesaw
\cite{minkowski1977} extension of the SM. The seesaw requires,
however, some generic lepton number violation as the neutrinos are
assumed to be Majorana particles. This lepton number violation can
arise from a spontaneous symmetry breaking in some more fundamental
theories. The simplest possibility is, for example, to consider the
singlet Majoron model \cite{cmp1980}, where the lepton number is a
global symmetry and its breaking will leave a massless
Nambu-Goldstone boson -- the Majoron. The global lepton number
breaking scale can be as low as the electroweak scale
\cite{jl1991,pilaftsis2008}. In this case the right-handed neutrinos
can be at an accessible scale which is is testable at colliders
\cite{hz2006}. At the same time the dark matter particle can have a
sizable coupling with the Majoron.

In this paper, we will study interplay of the dark-matter-Majoron
coupling on the dark-matter-Higgs coupling. For illustration, we
will focus on the simplest dark matter candidate, a real SM-singlet
scalar. This model proposed by Silveira and Zee \cite{sz1985} has
been studied before \cite{bpv2000,hllt2009,gms2010}. In these works,
dark matter annihilation is determined by the dark-matter-Higgs
coupling as the SM couplings are well known. Thus, direct detection
and relic density will both constrain the dark-matter-Higgs coupling
for a given dark matter mass. In the presence of a low lepton number
breaking scale, we can have a sizable dark-matter-Majoron coupling
besides an accessible seesaw scale. The dark matter thus could
significantly annihilate into the Majorons. In this case, the
dark-matter-Majoron coupling can affect the relic density in
addition to the dark-matter-Higgs coupling and the SM couplings. As
a result, a smaller dark-matter-Higgs coupling is required. This has
an interesting implication on dark matter direct detection
experiments.

\section{The Model}

In the singlet Majoron model \cite{cmp1980} \footnote{Alternatively,
one can also consider the spontaneous symmetry breaking of global
lepton number in other seesaw models with or without right-handed
neutrinos \cite{mw1980,zee1980,zee1985,ma1998,knt2003}.}, the
right-handed neutrinos $N_R^{}(\textbf{1},\textbf{1},0)$ have no
Majorana masses, which explicitly break the lepton number. Instead,
they have the following Yukawa couplings with a complex singlet
scalar $\sigma(\textbf{1},\textbf{1},0)$,
\begin{eqnarray}
\label{yukawa} \mathcal{L}\supset-\frac{1}{2}f
\sigma\overline{N}_R^{c}N_R^{}+\textrm{H.c.}\,,
\end{eqnarray}
which exactly conserves the lepton number. After the complex singlet
$\sigma$ develops a vacuum expectation value (VEV) to spontaneously
break the lepton number, the right-handed neutrinos $N_R^{}$ can
obtain their Majorana masses,
\begin{eqnarray}
\label{massr}
\mathcal{L}\supset-\frac{1}{2}m_N^{}\overline{N}_R^{c}N_R^{}+\textrm{H.c.}~~\textrm{with}~~m_N^{}=f\langle\sigma\rangle\,.
\end{eqnarray}
Consequently the seesaw for generating the small neutrino masses is
available as the right-handed neutrinos also have the Yukawa
couplings with the SM lepton and Higgs doublets,
\begin{eqnarray}
\label{yukawa2} \mathcal{L}\supset-y_\nu^{}\bar{l}_L^{}\tilde{\phi}
N_R^{}+\textrm{H.c.}\,.
\end{eqnarray}
Here $l_L^{}(\textbf{1},\textbf{2},-\frac{1}{2})$ and
$\phi(\textbf{1},\textbf{2},\frac{1}{2})$ denote the SM lepton and
Higgs doublets, respectively.

We now extend the singlet Majoron model with a real singlet scalar
$\chi(\textbf{1},\textbf{1},0)$. The full scalar potential should be
\begin{eqnarray}
V&=& \frac{1}{2}\mu_\chi^2 \chi^2_{}  +
\mu_\sigma^{}|\sigma|^2_{}+\mu_\phi^2|\phi|^2_{}+
\frac{1}{4}\lambda_\chi^{}\chi^4_{}+\lambda_\sigma^{}|\sigma|^4_{}\nonumber\\
&&+\lambda_\phi^{}|\phi|^4_{}+\alpha
\chi^2_{}|\sigma|^2_{}+\beta\chi^2_{}|\phi|^2_{}+\gamma|\sigma|^2_{}|\phi|^2_{}\,.
\end{eqnarray}
Here we have imposed a $Z_2^{}$ discrete symmetry under which only
the real singlet $\chi$ is odd while other SM and non-SM fields all
carry an even parity. The $Z_2^{}$ symmetry is required to hold at
any energy scales. So, the real singlet $\chi$ should be stable.

After the global and gauge symmetry breaking, we can describe the
singlet $\sigma$ and the doublet $\phi$ by
\begin{eqnarray}
\displaystyle{\sigma=\frac{1}{\sqrt{2}}(v'+h')e^{i\frac{\rho}{v'}}_{}}\,,~~\phi
= \left[
\begin{array}{c}
0\\
[2mm] \frac{1}{\sqrt{2}}(v+h)\end{array} \right]\,,
\end{eqnarray}
where $\rho$ is the massless Majoron whereas $v'$ and $v$ are the
vacuum expectation value (VEV),
\begin{eqnarray}
\label{vevs}
v'=\sqrt{\frac{-\mu_\sigma^2+\frac{\gamma}{2\lambda_\phi^{}}}{\lambda_\sigma^{}-\frac{\gamma^2_{}}{4\lambda_\phi^{}}}}\,,~~
v=\sqrt{\frac{-\mu_\phi^2+\frac{\gamma}{2\lambda_\sigma^{}}}{\lambda_\phi^{}-\frac{\gamma^2_{}}{4\lambda_\sigma^{}}}}\simeq
246\,\textrm{GeV} \,.
\end{eqnarray}
We thus can obtain the following masses and interactions of the
physical bosons,
\begin{eqnarray}
\label{potential2} V&\supset& \frac{1}{2}m_\chi^{} \chi^2_{}
+\frac{1}{2}m_{h'}^2 h'^2_{}+ \frac{1}{2}m_h^2 h^2_{}+m_{h'h}^2
h'h\nonumber\\
&&+\alpha v' h' \chi^2_{}+\beta v h \chi^2_{}\,.
\end{eqnarray}
Here the masses have been defined by
\begin{eqnarray}
&&m_\chi^2=\mu_\chi^2+\alpha^{}v'^2_{}+\beta v^2_{}\,,~~
m_{h'}^2=2\lambda_\sigma^{}v'^2_{}\,,\nonumber\\
&&m_h^2=2\lambda_\phi^{}v^2_{}\,,~~m_{h'h}^2=\gamma v'v\,.
\end{eqnarray}
The Higgs bosons $h'$ and $h$ now mix together. The mass eigenstates
should be given by
\begin{eqnarray}
h_1^{}=h'\cos\vartheta-h\sin\vartheta\,,~~
h_2^{}=h'\sin\vartheta+h\cos\vartheta\,,
\end{eqnarray}
with
\begin{eqnarray}
\vartheta=\frac{1}{2}\arctan\frac{\gamma v' v}{\lambda_\sigma^{}
v'^2_{}-\lambda_\phi^{} v^2_{}} \,,
\end{eqnarray}
and
\begin{subequations}
\begin{eqnarray}
m_{h_1^{}}^2=\lambda_\sigma^{}v'^2_{}+\lambda_\phi^{}v^2_{}+\sqrt{(\lambda_\sigma^{}v'^2_{}-\lambda_\phi^{}v^2_{})^2_{}+\gamma^2_{}v'^2_{}v^2_{}}\,,&&
\nonumber\\
&&\\
m_{h_2^{}}^2=\lambda_\sigma^{}v'^2_{}+\lambda_\phi^{}v^2_{}-\sqrt{(\lambda_\sigma^{}v'^2_{}-\lambda_\phi^{}v^2_{})^2_{}+\gamma^2_{}v'^2_{}v^2_{}}\,.&&
\nonumber\\
&&
\end{eqnarray}
\end{subequations}
From the kinetic term, it is also easy to derive the trilinear
coupling of the non-SM Higgs boson $h'$ to the Majoron $\rho$,
\begin{eqnarray}
\mathcal{L}_K^{}\supset(\partial_\mu^{}\sigma)^\ast_{}(\partial^\mu_{}\sigma)\Rightarrow
\frac{1}{v'}h'\partial_\mu^{}\rho \partial^\mu_{}\rho\,.
\end{eqnarray}

The existence of the massless Majoron will result in some
phenomenological implications. For example, at one-loop order the
right-handed neutrinos will mediate the lepton flavor violating
decays including $\mu\rightarrow \rho e$, $\tau\rightarrow \rho e$,
and $\tau\rightarrow \rho \mu$. The Majoron will also have
implications on astrophysics, such as the cooling rates of white
dwafts, the helium ignition process in red giants, and the energy
emission of neutron stars. Furthermore, the Majoron will contribute
to the relativistic degrees of freedom which has been stringently
constrained by Primordial Big-Bang Nucleosynthesis (BBN). After
taking all of the experimental limits into account, we can still
expect the global symmetry of lepton number to spontaneously break
near the electroweak scale \cite{jl1991,pilaftsis2008}. This is also
consistent with the stability and triviality bounds \cite{jl1991}.
For such a lepton number breaking scale, the seesaw can be detected
at colliders \cite{hz2006}.

In the following, we shall simply assume the right-handed neutrinos
$N_R^{}$ are heavier than the stable scalar $\chi$. Therefore the
right-handed neutrinos can decouple from the discussions on the dark
matter property.

\section{Dark-Matter-Nucleon Scattering}

The real SM-singlet $\chi$ has a trilinear coupling with the SM
Higgs boson $h$, see Eq. (\ref{potential2}). The t-channel exchange of $h$ will result in an
elastic scattering of $\chi$ by nucleons and hence a nuclear recoil.
The spin-independent cross section of the elastic scattering would
be
\begin{eqnarray}
\label{crosssectiondn} \sigma_{\chi N\rightarrow \chi N }^{}
&=&\frac{1}{\pi}\left[\left(-\frac{v'}{2v}\alpha\sin2\vartheta+\beta\sin^2_{}\vartheta\right)\frac{1}{m_{h_2^{}}^2}\right.\nonumber\\
&&
\left.+\left(\frac{v'}{2v}\alpha\sin2\vartheta+\beta\cos^2_{}\vartheta\right)\frac{1}{m_{h_1^{}}^2}\right]^2_{}\nonumber\\
&&\times\frac{\mu_r^2}{m_\chi^2}f^2_{}m_N^2 \,,
\end{eqnarray}
where $m_N^{}$ is the nucleon mass, $\mu_{r}^{}=m_\chi^{}
m_N^{}/(m_\chi^{} + m_N^{})$ is the reduced mass, the factor $f$ in
the range $0.14 <f<0.66$ with a central value $f=0.30$
\cite{aht2008} parameterizes the Higgs to nucleons coupling from the
trace anomaly, $\displaystyle{fm_{N}^{}\equiv\langle
N|\sum_q^{}m_q^{}\bar{q}q|N\rangle}$. With a small mixing angle
$\vartheta$, which can be naturally achieved for
$v'=\mathcal{O}(100\,\textrm{GeV}-1\,\textrm{TeV})$ and
$\gamma=\mathcal{O}(0.1-1)$, we can approximate $h_2^{}$ to be the
SM Higgs boson $h$ and then simplify the above formula,
\begin{eqnarray}
\label{crosssectiondn2} \sigma_{\chi N\rightarrow \chi N }^{}
=\frac{\beta^2_{}}{\pi}\frac{\mu_r^2}{m_\chi^2 m_h^4}f^2_{}m_N^2 \,.
\end{eqnarray}
In the following, we shall focus on this simplified case. If $\chi$
is the dark matter particle, the scattering cross section
(\ref{crosssectiondn}) should be stringently constrained by the dark
matter direct detection experiments. For example, we can obtain
\begin{eqnarray}
\sigma_{\chi N\rightarrow \chi N }^{} &=&3.8\times
10^{-44}_{}\,\textrm{cm}^2_{}\,\left(\frac{\beta}{0.06}\right)^2_{}\left(\frac{70\,\textrm{Gev}}{m_\chi^{}}\right)^2_{}\nonumber\\
&&\times
\left(\frac{120\,\textrm{Gev}}{m_h^{}}\right)^4_{}\left(\frac{f}{0.3}\right)^2_{}\,,
\end{eqnarray}
which is consistent with bound from the recent CDMS II result
\cite{ahmed2009}.

\section{Dark Matter Annihilation}

The real SM-singlet $\chi$ could provide the dark matter relic
density if its annihilation decouples at an appropriate freeze-out
temperature, which is determined by the thermally averaging cross
section \cite{kw1980,kt1990},
\begin{eqnarray}
\label{crosssectionthermal}
\langle\sigma_{\textrm{A}}^{}v_{\textrm{rel}}^{}\rangle
=\frac{\displaystyle{\int_{4m_\chi^2}^{\infty}s\sqrt{s-4m_\chi^2}K_{1}^{}\left(\frac{\sqrt{s}}{T}\right)
\sigma_{\textrm{A}}^{}v_{\textrm{rel}}^{}
ds}}{\displaystyle{\int_{4m_\chi^2}^{\infty}s\sqrt{s-4m_\chi^2}K_{1}^{}\left(\frac{\sqrt{s}}{T}\right)
ds}}\,,
\end{eqnarray}
where
\begin{eqnarray}
v_{\textrm{rel}}^{}=2\left(1-\frac{4m_\chi^2}{s}\right)^{\frac{1}{2}}_{}\,,
\end{eqnarray}
is the relative velocity with $s$ being the squared center of mass
energy. The total cross section
$\sigma_{\textrm{A}}^{}v_{\textrm{rel}}^{}$ could be conveniently
divided into two parts,
\begin{eqnarray}
\sigma_{\textrm{A}}^{}v_{\textrm{rel}}^{}=\sigma_{\chi\chi\rightarrow
\bar{f}f}^{} v_{\textrm{rel}}^{} +\sigma_{\chi\chi\rightarrow
\rho\rho}^{} v_{\textrm{rel}}^{}
\end{eqnarray}
with $f$ being the SM fermions. Here we have assumed the SM gauge
and Higgs bosons and the other fields for the seesaw are heavier
than $\chi$ so that these heavy fields will only give a negligible
contribution to the thermal averaging cross section.

We calculate
\begin{eqnarray}
\label{dmff} \sigma_{\chi\chi\rightarrow \bar{f}f}^{}
v_{\textrm{rel}}^{}
&=&\frac{\beta^2_{}}{\pi}\frac{1}{(s-m_h^2)^2_{}+m_h^2\Gamma_h^2}\nonumber\\
&&\times\sum_f^{} N_f^c
m_f^2\left(1-\frac{4m_f^2}{s}\right)^{\frac{3}{2}}_{}\,,\\
\label{dmmajoron} \sigma_{\chi\chi\rightarrow \rho\rho}^{}
v_{\textrm{rel}}^{}
&=&\frac{\alpha^2_{}}{8\pi}\frac{s}{(s-m_{h'}^2)^2_{}+m_{h'}^2\Gamma_{h'}^2}\,.
\end{eqnarray}
Here $N_f^{}=1$ for the leptons while $N_f^{}=3$ for the quarks. For
$m_b^{}\ll m_\chi^{}<m_W^{}$, we can take $s=4m_\chi^2$ to read
\begin{eqnarray}
\label{thermal}
\langle\sigma_{\textrm{A}}^{}v_{\textrm{rel}}^{}\rangle
&=&\langle\sigma_{\chi\chi\rightarrow \bar{f}f}^{}
v_{\textrm{rel}}^{}\rangle+\langle\sigma_{\chi\chi\rightarrow \rho\rho}^{} v_{\textrm{rel}}^{}\rangle\nonumber\\
&=&\frac{3\beta^2_{}}{\pi}\frac{m_b^2}{(4m_\chi^2-m_h^2)^2_{}}+\frac{\alpha^2_{}}{2\pi}\frac{m_\chi^2}{(4m_\chi^2-m_{h'}^2)^2_{}}\nonumber\\
&=&\sigma_0^{}\,.
\end{eqnarray}
By analytically solving the Boltzmann equations \cite{kt1990}, we
can determine the frozen temperature,
\begin{eqnarray}
x_f^{}=\frac{m_\chi^{}}{T_f^{}}&\simeq&\ln[0.038 g_\ast^{-1/2}
m_{\textrm{Pl}}^{}m_{\chi}^{}
\sigma_0^{}]\nonumber\\
&&-0.5\ln\{\ln[0.038 g_\ast^{-1/2} m_{\textrm{Pl}}^{}m_{\chi}^{}
\sigma_0^{}]\}\,,
\end{eqnarray}
and then the relic density
\begin{eqnarray}
\Omega_{\chi}^{}h^2_{}= 1.07\times
10^9_{}\frac{x_f^{}\,\textrm{GeV}^{-1}_{}}{(g_{\ast
S}^{}/\sqrt{g_\ast^{}})m_\textrm{Pl}^{}\sigma_0^{}}
\end{eqnarray}
Here $m_{\textrm{Pl}}^{}\simeq 1.22\times 10^{19}_{}\,\textrm{GeV}$
is the Planck mass and $ g_{\ast S}^{}\simeq g_{\ast}^{}\simeq 100$
is the relativistic degrees of freedom. For
$m_\chi^{}=70\,\textrm{GeV}$, we need
\begin{eqnarray}
\sigma_0^{}=1.47\times 10^{-9}_{}\,\textrm{GeV}^{-2}_{}
~~\textrm{and~then} ~~ x_f^{}=18.5
\end{eqnarray}
to generate the desired relic density,
\begin{eqnarray}
\Omega_{\chi}^{}h^2_{}=0.11\,.
\end{eqnarray}

If the dark-matter-Majoron coupling is absent, the thermal cross
section is only related to the dark-matter-Higgs coupling. For
$m_h^{}=120\,\textrm{GeV}$ and $m_\chi^{}=70\,\textrm{GeV}$, we can
determine $\beta=0.0486$ for generating a right relic density. The
induced dark-matter-nucleon scattering cross section is slightly
smaller than the experimental bound. Now the dark-matter-Majoron
coupling can significantly contribute to the dark matter
annihilation. For example, if we take $m_\chi^{}=70\,\textrm{GeV}$,
the annihilation of the dark matter into the Majorons can account
for the dark matter relic density when $m_{h'}^{}$ decreases from
$865\,\textrm{GeV}$ to $120\,\textrm{GeV}$ while $\alpha$ decreases
from $1$ to $0.00714$. Therefore, to make the dark matter
annihilation not too fast, the dark-matter-Higgs coupling must be
reduced in the presence of a significant annihilation of the dark
matter into the Majorons. In consequence, we will get a smaller
dark-matter-nucleon scattering cross section.

\section{Summary}

In some interesting dark matter models, the SM Higgs boson could be
an unique messenger between dark and visible matters. In such
models, usually the dark-matter-Higgs coupling fully determines the
dark matter relic density as the SM couplings are well known. The
dark-matter-Higgs coupling also opens the way for dark matter direct
detection experiments. So, the relic density and direct detection
will both constrain the dark-matter-Higgs coupling for a given dark
matter mass. We considered the singlet Majoron model \cite{cmp1980},
where the right-handed neutrinos obtain their Majorana masses after
the global lepton number is spontaneously broken. Associated with
the global symmetry breaking there will be a massless Majoron. If
the seesaw is expected to detect at colliders, the symmetry breaking
scale could not be much heavier than the electroweak scale. In this
case the dark matter particle could sizably couple to and hence
significantly annihilate into the Majoron. This implies a smaller
dark-matter-Higgs coupling and hence a smaller dark-matter-nucleon
scattering cross section. We demonstrated this possibility in the
simplest dark matter model where a real SM-singlet scalar acts as
the dark matter. Our conclusion could be applied to other dark
matter models (for instance, see \cite{hambye2009,ghsz2009}).

\vspace{5mm}

\textbf{Acknowledgement}: AA and ML are supported by the
Sonderforschungsbereich TR 27 of the Deutsche
Forschungsgemeinschaft. PHG is supported by the Alexander von
Humboldt Foundation.


\begin{thebibliography}{99}



\bibitem{dunkley2008}
J. Dunkley {\it et al.}, [WMAP Collaboration], Astrophys. J. Suppl.
\textbf{180}, 306 (2009).


\bibitem{sz1985}
V. Silveira and A. Zee, Phys. Lett. B \textbf{161}, 136 (1985).



\bibitem{bpv2000}
J. McDonald, Phys. Rev. D \textbf{50}, 3637 (1994); C.P. Burgess, M.
Pospelov, and T. ter Veldhuis, Nucl. Phys. B \textbf{619}, 709
(2001).



\bibitem{hllt2009}
X.G. He, T. Li, X.Q. Li, and H.C. Tsai, Phys. Lett. B \textbf{688},
332 (2010); M. Farina, D. Pappadopulo, and A. Strumia, Phys. Lett. B
\textbf{688}, 329 (2010); M. Asano and R. Kitano, Phys. Rev. D
\textbf{81}, 054506 (2010); W.L. Guo and Y.L. Wu, arXiv:1006.2518
[hep-ph].



\bibitem{gms2010}
P.H. Gu, E. Ma, and U. Sarkar, Phys. Lett.  B \textbf{690}, 145
(2010).

\bibitem{DMreviews}
F. Petriello and K.M. Zurek, JHEP \textbf{0809}, 47 (2008); J. Kopp,
T. Schwetz, and J. Zupan, JCAP \textbf{1002}, 014 (2010); and
references therein.


\bibitem{hambye2009}
T. Hambye, JHEP \textbf{0901}, 028 (2009).




\bibitem{stv2008}
T. Schwetz, M.A. T\'{o}rtola, and J.W.F. Valle, New J. Phys.
\textbf{10}, 113011 (2008).



\bibitem{minkowski1977}
P. Minkowski, Phys. Lett. B \textbf{67}, 421 (1977); T. Yanagida, in
{\it Proc. of the Workshop on Unified Theory and the Baryon Number
of the Universe}, ed. O. Sawada and A. Sugamoto (KEK, Tsukuba,
1979), p. 95; M. Gell-Mann, P. Ramond, and R. Slansky, in {\it
Supergravity}, ed. F. van Nieuwenhuizen and D. Freedman (North
Holland, Amsterdam, 1979), p. 315; S.L. Glashow, in {\it Quarks and
Leptons}, ed. M. L\'{e}vy {\it et al.} (Plenum, New York, 1980), p.
707; R.N. Mohapatra and G. Senjanovi\'{c}, Phys. Rev. Lett.
\textbf{44}, 912 (1980).


\bibitem{cmp1980}
Y. Chikashige, R.N. Mohapatra, and R.D. Peccei, Phys. Lett. B
\textbf{98}, 265 (1981).



\bibitem{jl1991}
G. Jungman and M.A. Luty, Nucl. Phys. B \textbf{361}, 24 (1991); A.
Dedes, T. Figy, S. Hoche, F. Krauss, and T.E.J. Underwood, JHEP
\textbf{0811}, 036 (2008); and references therein.

\bibitem{pilaftsis2008}
A. Pilaftsis, Phys. Rev. D \textbf{78}, 013008 (2008); and
references therein.


\bibitem{hz2006}
T. Han and B. Zhang, Phys. Rev. Lett. \textbf{97}, 171804 (2006); A.
Atre, T. Han, S. Pascoli, and B. Zhang, JHEP \textbf{0905}, 030
(2009).

\bibitem{mw1980}
M. Magg and C. Wetterich, Phys. Lett. B \textbf{94}, 61 (1980); J.
Schechter and J.W.F. Valle, Phys. Rev. D \textbf{22}, 2227 (1980);
T.P. Cheng and L.F. Li, Phys. Rev. D \textbf{22}, 2860 (1980); G.
Lazarides, Q. Shafi, and C. Wetterich, Nucl. Phys. B \textbf{181},
287 (1981); R.N. Mohapatra and G. Senjanovi\'{c}, Phys. Rev. D
\textbf{23}, 165 (1981).



\bibitem{zee1980}
A. Zee, Phys. Lett. B \textbf{93}, 389 (1980).

\bibitem{zee1985}
A. Zee, Phys. Lett. B \textbf{161}, 141 (1985); K.S. Babu, Phys.
Lett. B \textbf{203}, 132 (1988).





\bibitem{ma1998}
E. Ma, Phys. Rev. Lett. \textbf{81}, 1171 (1998); E. Ma, Phys. Rev.
D \textbf{73}, 077301 (2006).



\bibitem{knt2003}
L.M. Krauss, S. Nasri, and M. Trodden, Phys. Rev. D \textbf{67},
085002 (2003); K. Cheung and O. Seto, Phys. Rev. D \textbf{69},
113009 (2004).





\bibitem{aht2008}
S. Andreas, T. Hambye, and M.H.G. Tytgat, JCAP \textbf{0810}, 034
(2008).

\bibitem{ahmed2009}
Z. Ahmed {\it et al.}, [CDMS Collaboration], arXiv:0912.3592
[astro-ph.CO].


\bibitem{kw1980}
E.W. Kolb and S. Wolfram, Nucl. Phys. B \textbf{172}, 224 (1980).

\bibitem{kt1990}
E.W. Kolb and M.S. Turner, \textit{The Early Universe},
Addison-Wesley, 1990.




\bibitem{ghsz2009}
P.H. Gu, H.J. He, U. Sarkar, and X. Zhang, Phys. Rev. D \textbf{80},
053004 (2009).







\end{thebibliography}
\end{document}